# Low Profile Metamaterial Band-Pass Filter Loaded with 4-Turn Complementary Spiral Resonator for WPT Applications

Rasool Keshavarz and Negin Shariati
School of Electrical and Data Engineering
University of Technology Sydney
Sydney, Australia
Email: Rasool.Keshavarz@uts.edu.au, Negin.Shariati@uts.edu.au

*Abstract*— In this paper, a very compact ($0.03\lambda_g \times 0.18\lambda_g$) and low insertion loss (<0.4 dB) metamaterial band-pass filter (MBPF) at the center frequency of $f_0 = 730$ MHz is proposed, based on the rectangular-shape 4-turn complementary spiral resonators (4-CSR). The proposed MBPF consists of an interdigital capacitor as a series capacitance in the top layer, leading to improve the stopband performance in the pass-band range of 700~760 MHz, which makes it suitable for wireless power transfer (WPT) systems by rejecting unwanted signals. In order to validate the performance of the proposed technique, the MBPF is fabricated on the RO-4003 substrate and great agreement is achieved between simulated and measured results. The stop-band attenuations of greater than 52 dB and 20 dB are obtained around the $0.8 \times f_{cl}$ (lower cutoff frequency) and $1.2 \times f_{cu}$ (upper cutoff frequency), respectively.

*Keywords*— *Band-pass filter, metamaterials (MTMs), complementary split-ring resonator (CSRR), microstrip filter, wireless power transfer (WPT).*

## I. INTRODUCTION

Wireless Power Transfer (WPT) and RF (Radio Frequency) energy harvesting are promising techniques to provide sustainable energy sources for future autonomous sensors and the Internet of Things (IoT) devices [1-4]. Rectification of electromagnetic waves to provide DC power through wireless transmission has been the subject of research for decades [5], [6]. However, the majority of literature on the receiver side has been focused on narrow bandwidth rectifiers [7]. Therefore, from a practical perspective, collecting RF signals over a moderate frequency band could increase the output of DC power [8].

Since band-pass filter (BPF) is a key component in the front-end chain of receivers [9], well-designed BPFs are necessary for the receiver section of WPT systems to suppress harmonics and unwanted signals generated by a rectification device [1], [6]. Moreover, WPT receivers require low loss, low cost, low weight, and compact microwave filters for an optimum operation [10]. To address this, several implementation techniques such as coupled-line, interdigital, surface acoustic wave (SAW) and defected ground structure (DGS) filters have been proposed [9], [11].

Also, there has been a growing interest in the field of metamaterial structures to design a variety of filters [12]. To name a few, split-ring resonators (SRRs) and complementary of SSRs (CSRRs) have been attracting great interest among researchers due to their potential applications in metamaterials synthesis with negative effective permeability [13]. The concept of complementary split-ring resonators (CSRRs) is introduced based on resonant elements as an alternative in the design of metamaterials, providing an effective negative permittivity, rather than permeability. The CSRRs structure, which can be defined as a negative image of the SRR in planar technology, exhibits an electromagnetic behavior that is almost dual of the SRR. Based on the architectures of SRR and CSRR, transmission lines offer the possibility to modify their phase and impedance, which make them ideal applicants not only for filtering but also for microwave devices, in general [14].

The resonant-type approach is a dedicated method in the implementation of composite right/left-handed transmission lines (CRLH TLs) [15] and extended CRLH TL (E-CRLH TL) as metamaterial transmission lines (MTM TL) [16]. The first MTM TL with SRR and CSRR structures appeared in [17] which involved a conventional coplanar waveguide loaded with SRRs etched on the bottom layer of the substrate. Moreover, due to small dimensions of SRR and CSRR resonators, the subsequent transmission lines can be very compact which makes them suitable for applications where size is an essential metric and miniaturization is required [13].

In this paper, a new 4-turn complementary spiral resonator (4-CSR) is introduced to design a miniaturized BPF. CSR is developed by a single strip rolled up to form a spiral slot, which can be applied in miniaturized filters by increasing electrical length. To demonstrate enhanced miniaturization of the proposed model, three types of MTM TL loaded by CSRs including CSRRs, CSR2s, and CSR3s were investigated [18]. The achieved results in this paper show that the miniaturization effect of CSR2 and CSR3 is much better than CSRR [18].

In the proposed metamaterial band-pass filter (MBPF), the MTM TL is loaded with 4-CSR structure, leading to a decrease in the physical length. Moreover, using an interdigital capacitor as a series capacitance in the MTM TL of MBPF, makes the lower cutoff frequency very sharp. According to the low profile and high sharpness of cutoff frequency, the proposed MBPF is suitable for WPT applications to reduce unwanted signals generated by rectifiers.

The organization of the paper is mentioned below. Section II describes the theory and design procedure of the proposed MBPF. Section III presents simulation results which are verified by measurements. Finally, Section IV concludes the paper.

## II. THEORY OF THE PROPOSED METAMATERIAL BAND-PASS FILTER

In recent years, several types of CSRR configurations are presented such as edge-coupled SRR (EC-SRR), broadside-coupled SRR (BC-SRR), nonbianisotropic SRR (NB-SRR), double-split SRR (2-SRR), and two-turn spiral resonator (2-SR) [18]. 2-SR configuration provides a strong magnetic dipole at the resonance frequency, hence being useful in metamaterial structures as the resonance frequency of the 2-SR is half the resonance frequency of EC-SRR and NB-SRR with the same size and shape. Furthermore, the resonance frequency behavior in SRR configurations is similar to their complement [18].

In this paper, a complement of 4-SR as 4-CSR is used, which is etched on the ground plane (bottom layer) in the proposed MBPF (Fig. 1.a). Using 4-CSR on the bottom layer leads to achieving resonance frequency (operational frequency band) with a more compact structure compared to other configurations (2-SR, EC-SRR, NB-SRR, etc.).

Moreover, in this work, a rectangular-shape 4-CSR is considered (Fig. 1.b). As coupling between 4-CSR resonator and the top layer highly depends on their average separation, a poor coupling is anticipated with circular 4-CSR [17]. Hence, circular-shape 4-CSR was not considered in the proposed metamaterial band-pass filter (MBPF). The equivalent circuit model of the proposed MBPF is presented in Fig. 1.c. In this circuit, $L_R^c$ and $C$ indicates the per-section inductance and capacitance of the line (top layer), $C_R^c$ and $L_L^c$ model the 4-CSR, and $C_L^c$ denotes the series interdigital capacitor.

In fact, the 4-CSRs should be relatively close to the top layer to create a valid equivalent circuit model in Fig. 1.c. Otherwise, a significant portion of the line would lie outside the region occupied by the 4-CSRs and the approximation of the coupling capacitance by the line capacitance would not be accurate.

In order to investigate transmission properties of the proposed MBPF, the series impedance and parallel admittance are considered and equal to:

$$Z = j(\omega L_R^c + \frac{1}{\omega C_L^c}), \quad Y = j\left(\frac{1}{\frac{1}{\omega C_R^c + 1/\omega L_L^c} + \frac{1}{\omega C}}\right) \quad (1)$$

MBPF supports propagating mode in the frequency interval where $\beta.l$ in the below equation is a real number [16]:

$$cos(\beta.l) = 1 + ZY/2 \quad (2)$$

where $l$ is the physical length of the unit cell and $\beta$ is the propagation constant of the MBPF. Equation (2) indicates that propagating mode occurs in the region between two frequencies; lower cutoff frequency $f_{cl}$ and upper cutoff frequency $f_{cu}$.

Due to using an interdigital capacitor as series capacitance in the proposed MBPF, the series impedance of the equivalent circuit model is dominated by $C_L^c$ and, hence, $L_R^c$ has been ignored in equation (1) [19]. Moreover, the cutoff frequencies take the simpler form [17]:

$$f_{cl} = \frac{1}{2\pi}\frac{1}{\sqrt{L_L^c(C_R^c + \frac{4}{\frac{1}{C_L^c} + \frac{4}{C}})}}, \quad f_{cu} = \frac{1}{2\pi\sqrt{C_R^c L_L^c}} \quad (3)$$

According to the equivalent circuit model in Fig. 1.c, there is a transmission zero frequency given by [18]:

$$f_z = \frac{1}{2\pi\sqrt{L_L^c(C + C_R^c)}} \quad (4)$$

Firstly, according to equation (3) and (4), as long as $C \leq 4C_L^c$, $f_z$ and $f_{cl}$ are very close, and hence, the MBPF shows a very sharp cutoff in the lower edge of the pass-band, regardless of the number of stages. Secondly, based on the equation (3), to design MBPF with a relatively wide pass-band, it is necessary to choose the high values of $C_L^c$ and $C$, as compared to the equivalent capacitance of the 4-CSR ($C_R^c$). Consequently, by using the interdigital capacitor as $C_L^c$ instead of a conventional gap, the value of this capacitance is increased and hence, a very sharp lower cutoff frequency is achieved and the bandwidth is improved.

## III. SIMULATION, MEASUREMENT, AND DISCUSSION

To confirm the design methodology, the MBPF prototype was fabricated on a Ro-4003 substrate with a thickness of 32 mils, $\varepsilon_r$=3.38 and tanδ=0.0035 (Fig. 2).

The dimension of the 4-CSR structure (Fig. 1.b) is presented in Table I. Table II shows the lumped components values in the equivalent circuit model (Fig. 1.c).

The MBPF has been simulated using Advanced Design System (ADS) and measured using a PNA Microwave Network Analyzer (N5245B). Fig. 3 compares simulated and measured results of the proposed MBPF. The measured pass-band is centered at 730 MHz, with a 3-dB bandwidth of 60 MHz and fractional bandwidth of 8.2%. The measured insertion loss including SMA connectors losses is 0.4 dB at the pass-band frequencies, and the return loss is better than 26 dB. Also, the stop-band attenuations >52 dB and >20 dB are obtained using the proposed technique around 0.8×$f_{cl}$ and 1.2×$f_{cu}$, respectively.

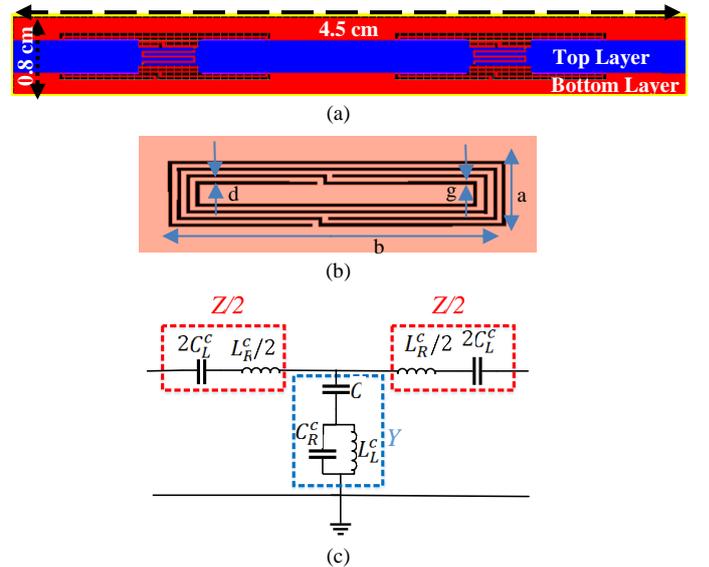

Fig. 1. a) The proposed MBPF Layout, b) 4-CSR structure, c) equivalent circuit model of the MBPF.

Table III compares the proposed filter characteristics with other published band-pass filters. It can be seen that the proposed MBPF achieved lower insertion loss, higher rejection of the lower cutoff frequency, and very small dimensions, compared to other works. It should be mentioned that a smaller 4-CSR with higher resonance frequency can be used on the bottom layer of the MBPF to improve the rejection of the upper cutoff frequency band but, the insertion loss increases, while the current upper band rejection (>20 dB) is sufficient for WPT applications [5].

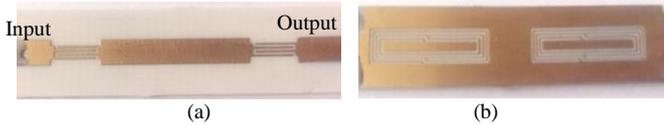

Fig. 2. Fabricated prototype of the proposed MBPF a) Top layer, b) Bottom layer.

Table I. DIMENSIONS OF THE PROPOSED 4-CSR.

| $a$ (mm) | $b$ (mm) | $d$ (mm) | $g$ (mm) |
|---|---|---|---|
| 3.5 | 15 | 0.2 | 0.2 |

Table II. LUMPED COMPONENTS VALUES IN THE EQUIVALENT CIRCUIT MODEL OF THE PROPOSED MBPF.

| $C_R^c$ | $L_L^c$ | $L_R^c$ | $C_L^c$ | $C$ |
|---|---|---|---|---|
| 4.1pF | 0.7 nH | 10.3nH | 0.3 pF | 2.1 pF |

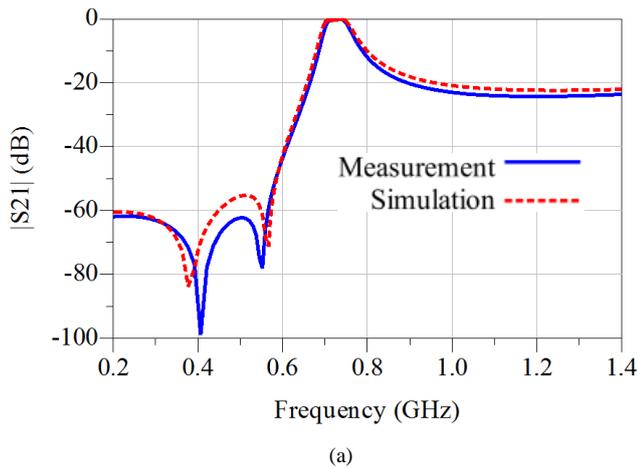

(a)

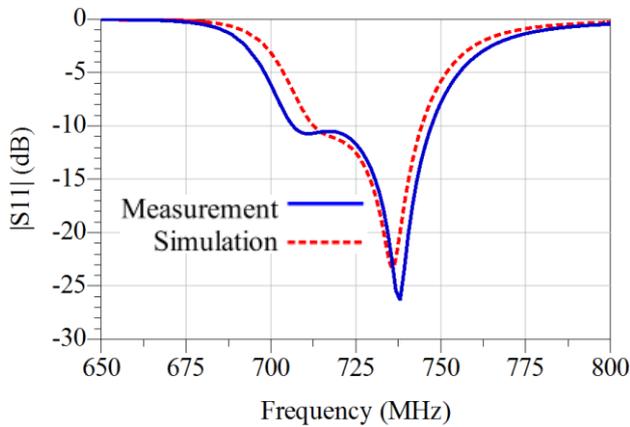

(b)

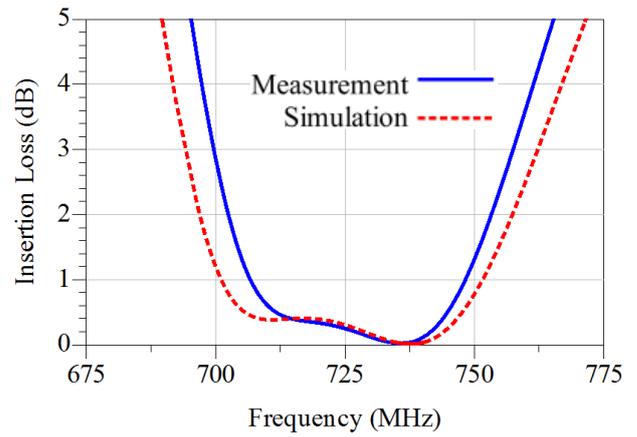

(c)

Fig. 3. Measured and simulation results of the proposed MBPF, a) |S21|, b) insertion loss, c) |S11|.

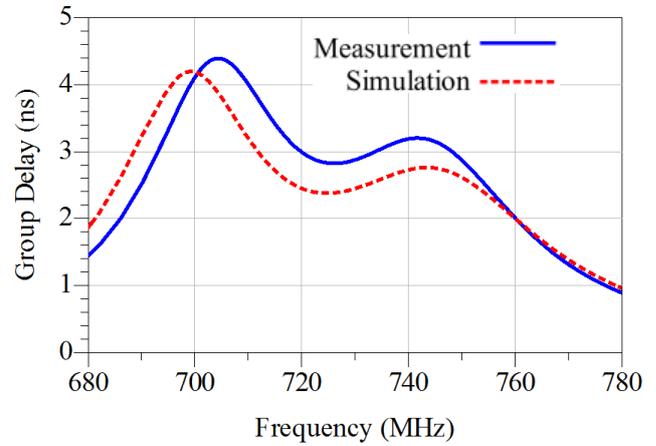

Fig. 4. Measured and simulation results of the group delay of the proposed MBPF.

IV. CONCLUSION

In this paper, a very compact MBPF has been proposed, which is loaded with 4-CSR structure. In addition, a high and broad stop-band suppression level is achieved in the lower cutoff frequency (>52 dB) using an interdigital capacitor as a series capacitance in the proposed MBPF. The compact planar MBPF, exhibits very low insertion loss (<0.4 dB) which makes it suitable for modern wireless communication systems and WPT applications.

Table III. Comparison of the Proposed MBPF With Other Works.

| Ref. | Center frequency ($f_0$) (GHz) | 3-dB Bandwidth (MHz) | Fractional Bandwidth (%) | Insertion Loss (dB) | Return Loss @$f_0$ (dB) | Stop-band Attenuation (dB) @$0.8f_{cl}$ | Stop-band Attenuation (dB) @$1.2f_{cu}$ | Dimensions ($\lambda_g^2$) |
|---|---|---|---|---|---|---|---|---|
| [12] | 3.3 | 50 | 1.5 | 1.6 | 19 | 20 | 30 | 0.4×0.4 |
| [20] | 6 | 1100 | 18 | 1.1 | 18 | 12 | 32 | 0.25×0.8 |
| [21] | 8.46 | 3430 | 40 | 1.5 | 25 | 10 | 30 | 0.6×0.7 |
| [22] | 10 | 1950 | 19 | 3.6 | 25 | 30 | 10 | 0.5×2 |
| [23] | 3.42 | 120 | 3.5 | 1 | 18 | 20 | 25 | 0.5×0.5 |
| [24] | 9.92 | 1010 | 10 | 1 | 15 | 30 | 45 | 0.5×0.5 |
| **Our work** | 0.73 | 60 | 8.2 | 0.4 | 26 | >52 | >20 | 0.03×0.18 |